\documentclass[aps]{revtex4}
\usepackage[colorlinks=true, pdfstartview=FitV, linkcolor=blue, citecolor=red, urlcolor=magenta, breaklinks=true]{hyperref}
\usepackage{graphicx}  
\usepackage{amsfonts}
\begin{document}
\title{Gravitational Aharonov-Bohm effect due to noncommutative BTZ black hole}
\author{M. A. Anacleto, F. A. Brito and E. Passos}
\email{anacleto, fabrito, passos@df.ufcg.edu.br}
\affiliation{Departamento de F\'{\i}sica, Universidade Federal de Campina Grande, Caixa Postal 10071, 58109-970 Campina Grande, Para\'{\i}ba, Brazil}
\begin{abstract} 
{ In this paper we consider the scattering of massless planar scalar waves by a noncommutative BTZ black hole. We compute the differential cross section via the partial wave approach, and 
we mainly show that the scattering of planar waves leads to a modified Aharonov-Bohm effect due to spacetime noncommutativity.}

\end{abstract}
\maketitle
\pretolerance10000

\section{Introduction}
 Noncommutative theories have been discussed in the literature by many authors~\cite{SW,SGhosh,rivelles,revnc}. The inherent nonlocality of these theories leads to the surprising mixture between ultraviolet (UV) and infrared (IR) divergences~\cite{SMRS} which could break the perturbative expansion, lead to loss of unitarity~\cite{GMehen} and violation of Lorentz invariance~\cite{Susskind}.
Furthermore, the study of noncommutative black holes have also been investigated by many authors in recent years. 
For example, the noncommutative Banados-Teitelboim-Zanelli (BTZ) {black holes were first analyzed in~\cite{Banados} and in~\cite{Chang} 
the noncommutative BTZ metric} was found based on the three dimensional equivalence between gravity and the Chern-Simons theory   which is a 3-dimensional topological quantum field theory and using the Seiberg-Witten map with the commutative BTZ solution~\cite{BTZ}.  
The BTZ black hole is a solution of (2+1) dimensional gravity with negative cosmological
constant and has become an important field of investigations.  
It is now well accepted that three-dimensional gravity is an excellent laboratory in order to explore and test some of the ideas behind the AdS/CFT correspondence~\cite{ADS}

Noncommutative field theories have also been object of several investigations in planar physics.
Among several topics in planar physics, the Aharonov-Bohm (AB) effect~\cite{Bohm} is one of the most extensively studied problems. This effect is essentially the scattering of charged particles
by a flux tube and has been  {experimentally confirmed by Tonomura~\cite{RGC} and for review see~\cite{Peskin}}. In  quantum field theory the effect has been simulated, for instance, by using a
nonrelativistic field theory describing bosonic particles
interacting through a Chern-Simons field~\cite{BL}.  It was also found to have analogues in several physical systems such as  in gravitation \cite{FV}, fluid dynamics
\cite{CL}, optics \cite{NNK} and Bose-Einstein condensates \cite{LO} appearing in a vast literature.

The noncommutative AB effect has been already studied in the context of quantum mechanics
\cite{FGLR,Chai} and in the quantum field theory approach~\cite{An}.
In~\cite{FGLR} the noncommutative AB effect has been
shown to be in contrast with the commutative situation. 
It was shown that the cross section for the scattering of scalar particles by a
thin solenoid does not vanish even if the magnetic
field assumes certain discrete values. 

{ Recently, it was shown in~\cite{Dolan} that the scattering of planar waves by a
draining bathtub vortex describes a modified AB effect
which depends on two dimensionless parameters associated with the circulation and draining rates \cite{Fetter}.  The
effect was shown to be inherently asymmetric even in the low-frequency limit and leads to novel interference patterns. 
In addition we consider the acoustic black hole metrics obtained from a relativistic fluid
in a noncommutative spacetime~\cite{ABP12} via the Seiberg-Witten map and also obtained from the Lorentz violating 
Abelian Higgs model~\cite{ABP11}. More recently in~\cite{ABP2012-1}, we have extended the analysis made in~\cite{Dolan}  to a Lorentz-violating and noncommutative background \cite{Bazeia:2005tb} which allows to have persistence of phase shifts even if circulation and draining vanish.}

 In general relativity  the study of the  absorption and scattering cross sections of planar waves in the vicinity  of black holes have been extensively studied in the literature by many authors and are of great relevance to experimental investigation~\cite{fhm}. With the discovery of a Higgs-like particle by the ATLAS and CMS collaborations has given extra motivation for studying absorption and scattering of bosonic fields with mass~\cite{ATLAS}.

In our study we apply the noncommutative BTZ metric to determine the differential cross section due to the scattering  of planar waves that leads to a modified AB effect in a noncommutative spacetime.  
We anticipate that we have obtained a cross section similar to that obtained  in~\cite{ABP2012-1} for an analogue Aharonov-Bohm effect due to an idealized draining bathtub vortex and in~\cite{FGLR} for noncommutative AB effect in quantum mechanics. 
The result implies that due to the spacetime noncommutativity pattern fringes
can still persist even in the limit where the parameters associated with the circulation go to zero (non-rotating black-holes). 
 In this limit, the noncommutative  background forms a conical defect, which is also responsible for the appearance of the analogue AB effect \cite{ABP2012-1}.
 
 The paper is organized as follows. In Sec.~\ref{II} we briefly introduce the noncommutative BTZ black hole background. In Sec.~\ref{AB-gravit} we shall compute 
 the differential cross section due to the scattering  of planar waves that leads to a modified AB effect in the noncommutative spacetime. Finally in Sec.~\ref{conclu} we present
 our final considerations.

\section{ Noncommutative BTZ black holes}
\label{II}
The metric of noncommutative BTZ black hole is given by~\cite{Chang}
\begin{eqnarray}
ds^2&=&-F^2dt^2+N^{-2} dr^2+2r^2N^{\phi}dtd\phi+\Big(r^2-\frac{\theta B}{2} \Big) d\phi^2 + {\cal O}(\theta^2),
\end{eqnarray}
where
\begin{eqnarray}
F^2&=&\frac{r^2-r^2_{+}-r^2_{-}}{l^2}-\frac{\theta B}{2},
\\
N^2&=&\frac{1}{r^2l^2}\left[(r^2-r^2_{+})(r^2-r^2_{-}) -\frac{\theta B}{2}(2r^2-r^2_{+}-r^2_{-})\right],
\\
N^{\phi}&=&\frac{-r_{+}r_{-}}{lr^2},
\end{eqnarray}
here $B$ is the magnitude of the magnetic field, $ \theta $ is the noncommutative parameter, $ r_{+} $ and $r_{-}$ are the outer and inner horizons of the commutative BTZ black hole given by
\begin{eqnarray}
r^2_{\pm}=\frac{l^2M}{2}\left[1\pm\sqrt{1-\left(\frac{J}{Ml}\right)^2}  \right].
\end{eqnarray}
The apparent horizons, for the noncommutative black hole, denoted by $ \hat{r}_{\pm} $ can be
determined by $ N^{2}=0 $
\begin{eqnarray}
\hat{r}_{\pm}=r_{\pm}+\frac{\theta B}{2} + {\cal O}(\theta^2),
\end{eqnarray}
note that the event horizons in the noncommutative case, are shifted through constant $ \theta B/2 $. In the limit $ \theta\rightarrow 0 $, these reduce to the event horizons of the commutative case.

The metric of noncommutative BTZ black hole can be rewritten as
\begin{eqnarray}
\label{mbtz}
ds^2&=&-fdt^2+{\cal Q}^{-1} dr^2-\frac{J}{r}rdtd\phi+\Big(1-\frac{\theta B}{2r^2} \Big) r^2d\phi^2.
\end{eqnarray}
where
\begin{eqnarray}
f&=&-M+\frac{r^2}{l^2}-\frac{\theta B}{2},
\\
{\cal Q}&=&-M+\frac{r^2}{l^2}+\frac{J^2}{4r^2}-\frac{\theta B}{2}\left( \frac{2}{l^2}-\frac{M}{r^2} \right),
\end{eqnarray}

The metric can be now written in the form
\begin{eqnarray}
g_{\mu\nu}=\left[\begin{array}{clcl}
-f &\quad\quad 0& -\frac{J}{2r}\\
0 & \quad\quad {\cal Q}^{-1}& 0\\
-\frac{J}{2r} &\quad\quad 0 & \left( 1-\frac{\theta B}{2r^2} \right)
\end{array}\right],
\end{eqnarray}
with inverse $g^{\mu\nu}$
\begin{eqnarray}
\label{metrinv}
g^{\mu\nu}=\frac{1}{-g}\left[\begin{array}{clcl}
-\Big(1-\frac{\theta B}{2r^2}\Big){\cal Q}^{-1} &\quad\quad 0& \quad\quad-\frac{J}{2r{\cal Q}}\\
0 & \quad -g{\cal Q}&\quad\quad 0\\
-\frac{J}{2r{\cal Q}} &\quad\quad 0 & \quad\quad \frac{f}{{\cal Q}}
\end{array}\right],
\end{eqnarray}
where $-g=\left[\frac{\theta B}{2}\left(-1-\frac{1}{l^2}+\frac{M}{r^2}\right) +\frac{J^2}{4r^2}+\frac{r^2}{l^2}-M \right]{\cal Q}^{-1}=\left[\frac{\theta B}{2}\left(\frac{1}{l^2}-1\right)+{\cal Q}\right]{\cal Q}^{-1}$. 

In order to study the Aharonov-Bohm effect, we shall now consider the Klein-Gordon equation in the noncommutative BTZ metric given by the background (\ref{mbtz}):
\begin{eqnarray}\label{KG-19-08}
\frac{1}{\sqrt{-g}}\partial_{\mu}(\sqrt{-g}g^{\mu\nu}\partial_{\nu})\psi=0.
\end{eqnarray}
As usual, one can make a separation of variables into the equation (\ref{KG-19-08}) as in the following
\begin{eqnarray}
\psi(t,r,\phi)=R(r)e^{i(\omega t-m\phi)}.
\end{eqnarray}
Let us now concentrate on the radial function $R(r)$ that satisfies the following linear second-order differential equation
\begin{eqnarray}
\label{EQKG}
&&\left[\Big(1-\frac{\theta B}{2r^2}\Big)\omega^2-\frac{Jm\omega}{r^2}-\frac{m^2f}{r^2}\right]\frac{R(r)}{(-g){\cal Q}}
+\frac{1}{r\sqrt{-g}}\frac{d}{dr}\left[r\sqrt{-g}{\cal Q}\frac{d}{dr}\right]R(r)=0.
\end{eqnarray}
The equation (\ref{EQKG}) can be rewritten as
\begin{eqnarray}
\label{EQKGbtz}
&&\left[\Big(1-\frac{\theta B}{2r^2}\Big)\omega^2-\frac{Jm\omega}{r^2}-\frac{m^2f}{r^2}\right]R(r)
+\frac{{\cal F}(r)}{r}\frac{d}{dr}\left[r{\cal F}(r)\frac{d}{dr}\right]R(r)=0,
\end{eqnarray}
where $ {\cal F}(r)=\sqrt{-g}{\cal Q}(r) $.  
{Now introducing the coordinate $\varrho$ through the use of the following equation~\cite{Gamboa}}
\begin{eqnarray}
\frac{d}{d{\varrho}}={\cal F}(r)\frac{d}{dr},
\end{eqnarray}
and considering a new radial function, $G(\varrho)=r^{1/2}R(r)$ we get to a new radial equation obtained from (\ref{EQKGbtz}) that reads
\begin{eqnarray}
\label{EG}
\frac{d^2G(\varrho)}{d\varrho}+\left\{\left[\left(1-\frac{\theta B}{4r^2}\right)\omega
-\frac{Jm}{2r^2}\left(1+\frac{\theta B}{4r^2}\right) \right]^2-V(r)\right\}G(\varrho)=0, 
\end{eqnarray}
where $V(r)$ is the potential given by
\begin{eqnarray}
V(r)=\frac{{\cal F}(r)}{4r^2}\left[\frac{4m^2f(r)}{{\cal F}(r)}-1
+\frac{J^2m^2}{r^2{\cal F}(r)}\left(1+\frac{\theta B}{4r^2}\right)+2r\frac{d{\cal F}(r)}{dr}\right],
\end{eqnarray}
a form that resembles that given in Refs. \cite{Dolan, ABP2012-1}. {Notice that this potential does not satisfy the asymptotic behavior $V(r)\rightarrow 0 $ as $ r\rightarrow\infty $. 

Now  we rewrite the equation (\ref{EG}) in terms of the new function $X(r)={\cal F}(r)^{1/2}G(\varrho)$, that is }
\begin{eqnarray}
\label{eqX}
\frac{d^2X(r)}{dr^{2}}&+& \left[-\frac{1}{2{\cal F}(r)}\frac{d^2{\cal F}(r)}{dr^2}+\frac{1}{4{\cal F}^2(r)}\left(\frac{d{\cal F}(r)}{dr} \right)^2\right]X(r)
\nonumber\\
&+&\left\{ \left[\left(1-\frac{\theta B}{4r^2}\right)\omega
-\frac{Jm}{2r^2}\left(1+\frac{\theta B}{4r^2}\right) \right]^2-V(r)\right\}
\frac{X(r)}{{\cal F}^2(r)}=0, 
\end{eqnarray}
where 
\begin{equation}
\frac{d{\cal F}(r)}{dr}=\frac{1}{{\cal F}}\left[{\cal Q}+\frac{\theta B}{4}\left(\frac{1}{l^2}-1 \right) \right]
\left[\frac{2r}{l^2}-\frac{(J^2+2\theta B)}{2r^3} \right],
\end{equation}
and 
\begin{eqnarray}
\frac{d{^2\cal F}(r)}{dr^2}&=&\frac{1}{{\cal F}}\left\{\left[ {\cal Q}+\frac{\theta B}{4}\left(\frac{1}{l^2}-1 \right) \right]
\left[\frac{2}{l^2}+\frac{3(J^2+2\theta B)}{4r^4} \right]+\left[\frac{2r}{l^2}-\frac{(J^2+2\theta B)}{2r^3} \right]^2\right\}
\nonumber\\
&-&\frac{1}{{\cal F}^3}\left[ {\cal Q}+\frac{\theta B}{4}\left(\frac{1}{l^2}-1 \right) \right]^2
\left[\frac{2r}{l^2}-\frac{(J^2+2\theta B)}{2r^3} \right]^2.
\end{eqnarray}
The equation (\ref{eqX}) can still be written in terms of a power series in $1/r$ as follows
\begin{eqnarray}
\label{eqv}
\frac{d^2X(r)}{dr^{2}}
+\left[-\frac{1}{r^2}+{\cal V}(r)+U(r)\right]X(r)=0, 
\end{eqnarray}
where
\begin{eqnarray}
{\cal V}(r)&=&\frac{1}{4\omega^2 r^4}\left[-72 \omega^2\Theta +b^2\left(1- 24 \Theta  -  4m^2 - 48M\right )+4b^4\right]
\nonumber\\
&+&\frac{1}{8\omega^4 r^6}\Big[b^2\Big(-8Mb^2\widetilde{m}^2  
+ 200 \Theta M \omega^2- 17 \Theta \omega^2 \Big)
+b^4\Big( 2 M+ 56 M^2 + 56 \Theta  M  + \Theta
+ 16 \Theta \omega^2\Big) 
 \nonumber\\
&+& b^6(16 M + 8 \Theta )\Big]
\\
U(r)&=&\frac{1}{32\omega^6r^8}\Big[ a^2 b^2(38 \Theta b^2  + 76 Mb^2 + 2b^2 + 
 114 \Theta \omega^2) -a b^4 m ( 64 b^2  M  +32 \Theta b^2   + 96 \Theta\omega^2)   
\nonumber\\ 
 &+& b^4 M\omega^2(96 \Theta m^2 -  320 \Theta M +  584 \Theta )+b^6 M(64 m^2 M + 96 \Theta m^2 - 16 M - 16 \Theta - 64 \Theta \omega^2)\Big]
\end{eqnarray}
 where $\widetilde{m}^2=m^2(1+\frac{3\omega^2\Theta}{Mb^2})+\frac{am}{M}-\frac{3a^2}{8Mb^2}$, $ \Theta=\theta B/2 $, $a={\omega}J$ and $ b=l\omega $. {Notice that the total potential $V(r)\equiv-1/r^2+{\cal V}(r)+U(r)$ in equation (\ref{eqv}) satisfies the expected asymptotic behavior $ V(r)\rightarrow 0 $ as $ r\rightarrow\infty $. Thus, we are now able to apply the usual techniques to look for Aharonov-Bohm effect. }
 
 \section{Gravitational Aharonov-Bohm effect}
\label{AB-gravit}
{To address the issues concerning the gravitational Aharonov-Bohm effect we shall now consider the scattering of a monochromatic planar wave of frequency $\omega$ written in the form \cite{jack}}
\begin{eqnarray}
\psi(t,r,\phi)=e^{-i\omega t}\sum_{m=-\infty}^{\infty}R_{m}(r) e^{im\phi}/\sqrt{r},
\end{eqnarray}
such a way that far from the vortex, the function $\psi$ can be given in terms of the sum of a plane wave and a scattered wave:
\begin{eqnarray}
\psi(t,r,\phi)\sim e^{-i\omega t}(e^{i\omega x}+f_{\omega}(\phi)e^{i\omega r}/\sqrt{r}),
\end{eqnarray}
where $e^{i\omega x}=\sum_{m=-\infty}^{\infty}i^mJ_{m}(\omega r) e^{im\phi}$ and $J_{m}(\omega r)$ 
is a Bessel function of the first kind. The scattering amplitude $f_{\omega}(\phi)$ has the following
partial-wave representation
\begin{eqnarray}
f_{\omega}(\phi)= \sqrt{\frac{1}{2i\pi\omega}}\sum_{m=-\infty}^{\infty}(e^{2i\delta_{m}}-1) e^{im\phi}.
\end{eqnarray}
 {Now to compute the phase shift $\delta_{m} $ we apply the folowing approximation formula}
\begin{eqnarray}
\delta_{m}\approx \pi (m-\tilde{m})+{\pi}\int^{\infty}_{0}r[J_{\tilde{m}}({\omega}r)]^2U(r)dr,
\end{eqnarray}
 and using $|m|\gg\sqrt{a^2+b^2}$, we obtain~\cite{Dolan,ABP2012-1}
\begin{eqnarray}
\label{fase}
\delta_{m}\cong {\pi}\left[1-\sqrt{1+\frac{3\omega^2\Theta}{Mb^2}}\right]m\frac{m}{|m|}
-\frac{a\pi}{2M\sqrt{1+\frac{3\omega^2\Theta}{Mb^2}}}\frac{m}{|m|}+O(m^{-1},a^2,b^2).
\end{eqnarray}
{ Therefore, we can compute the differential scattering cross section  
restricted to small angles $\phi$ and to lower orders in $\Theta$ that is given by 
}
\begin{eqnarray}
\label{sc}
\frac{d\sigma}{d\phi}&=&|f_{\omega}(\phi)|^2
\cong \frac{\pi^2 a^2}{2\pi\omega M^2}\left(1-\frac{3\omega^2\Theta}{b^2 M}+\frac{9\omega^4\Theta^2}{b^4 M^2}+O(\Theta^3) \right)\left[\left(\frac{4}{\phi^2}-\frac{2}{3}+\frac{\phi^2}{60}+O(\phi^3)\right) \right.
\nonumber\\
&+&\left.\left(- \frac{8\pi}{\phi^3}+\frac{\pi\phi}{30}+O(\phi^3) \right)\left(-\frac{3\omega^2\Theta}{2b^2M} +\frac{9\omega^4\Theta^2}{8b^4M^2}+O(\Theta^3) \right)\right.
\nonumber\\
&+&\left.\left(\frac{9\omega^4\Theta^2}{4 b^4 M^2}+O(\Theta^3) \right)\left(\frac{16\pi^2}{\phi^4}-\frac{4\pi^2}{3\phi^2}-\frac{4\pi^2}{45}-\frac{\pi^2 \phi^2}{3780}+O(\phi^3)\right)\right]
\nonumber\\
&+&\frac{1}{2\pi\omega}\left(\frac{9\omega^4\Theta^2}{4 b^4 M^2}+O(\Theta^3) \right)\left(\frac{16\pi^2}{\phi^4}+\frac{8\pi^2}{3\phi^2}+\frac{11\pi^2}{45}+\frac{31\pi^2 \phi^2}{1890}+O(\phi^3)\right).
\end{eqnarray}
Note that if $\Theta=0 $ in (\ref{fase}), we simply have 
\begin{eqnarray}
\delta_{m}\cong -\frac{a\pi}{2M}\frac{m}{|m|}.
\end{eqnarray}
This phase shift leads to a gravitational AB effect and the differential scattering cross section (\ref{sc}) at small angles $\phi$ is given by
\begin{eqnarray}
\frac{d\sigma}{d\phi}=|f_{\omega}(\phi)|^2\cong\frac{\pi^2 a^2}{2\pi\omega M^2}\left[\frac{4}{\phi^2}-\frac{2}{3}+\frac{\phi^2}{60}+O(\phi^3)\right],
\end{eqnarray}
{ whose leading term of the differential cross section reads}
\begin{eqnarray}
\frac{d\sigma}{d\phi}=\frac{\pi^2 \tilde{a}^2}{2\pi\omega}\frac{4}{\phi^2},
\end{eqnarray}
where, $ \tilde{a}=a/M $. This is the differential scattering cross section for the gravitational AB effect.

On the other hand, for $a=0$ 
and restricting  ourselves to lower orders in $\Theta$, the differential scattering cross section (\ref{sc}) now becomes
\begin{eqnarray}
\frac{d\sigma}{d\phi}=|f_{\omega}(\phi)|^2\cong\frac{1}{2\pi\omega}\left[\frac{9\omega^4\Theta^2}{4 b^4 M^2}+O(\Theta^3) \right]\left(\frac{16\pi^2}{\phi^4}+\frac{8\pi^2}{3\phi^2}+\frac{11\pi^2}{45}+\frac{31\pi^2 \phi^2}{1890}+O(\phi^3)\right).
\end{eqnarray}
{ Since we are taking $\phi \rightarrow 0 $, the differential cross section is dominated by}
\begin{eqnarray}
\frac{d\sigma}{d\phi}=\frac{9\Theta^2 }{2\pi\omega l^4 M^2}\frac{4\pi^2}{\phi^4}.
\end{eqnarray}
Note that the noncommutative gravitational AB scattering result through the use of partial waves approach is successfully
obtained  and contrarily to the usual Aharonov-Bohm effect, in the noncommutative case the differential
scattering cross section is different from zero even if $a=0$. 
{Moreover, for $ \Theta=0 $ and $ a\neq 0 $ in (\ref{sc})  the scattering cross section is symmetric under $ \phi \rightarrow-\phi $, 
while for $\Theta\neq 0 $ and $ a\neq 0 $ the low-frequency scattering cross section becomes asymmetric, with 
\begin{eqnarray}
\sigma_\bot=\int_{-\pi}^{\pi}\frac{d\sigma}{d\phi}\sin\phi d\phi
=\frac{3\pi^3 a^2\omega\Theta}{ b^2M^3}+O(\Theta^2)=\frac{3\pi^3 a^2\Theta}{ \omega l^2M^3}+O(\Theta^2).
\end{eqnarray}
This result is due to contribution $ 1/\phi^3 $ in the differential cross section.
}

\section{Conclusions}
\label{conclu}

In summary, in this paper we have considered the gravitational Aharonov-Bohm effect in the background of the noncommutative BTZ black hole. To address the issues concerning the noncommutative gravitational Aharonov-Bohm effect we considered the scattering of a monochromatic planar wave.
Our results are qualitatively in agreement with that obtained in~\cite{FGLR} for the AB effect in the context of noncommutative quantum mechanics and in~\cite{ABP2012-1} for an analogue Aharonov-Bohm
effect due to an idealized draining bathtub vortex. The noncommutative correction vanishes in the limit $\Theta\rightarrow 0$ so that no singularities are generated.
The leading correction ($\sim\Theta^2$) due to effect of spacetime noncommutativity may be relevant at a scale where the spacetime noncommutativity takes place as expected in high energy physics.
Our result shows that pattern fringes can appear even if $a$, which depends on the angular momentum, is equal to zero, unlike the commutative case. One can make some estimative of the aforementioned effect by estimating  $\theta$ following similar calculations already known in the literature \cite{casana,mckellar}.

\acknowledgments

We would like to thank CNPq, CAPES and PNPD/PROCAD -
CAPES for partial financial support.

\end{document}